\documentclass[aps,prd,showpacs,floatfix,nofootinbib,twocolumn,10pt]{revtex4}

\usepackage{color,amsmath,amssymb,graphicx,latexsym,subfigure}
\usepackage{threeparttable,multirow,txfonts,lineno}

\begin{document}

\title{Implication of multiple source populations of Galactic cosmic rays 
from proton and helium spectra}

\author{Qiang Yuan$^{a,b}$\footnote{yuanq@pmo.ac.cn}}

\affiliation{
$^a$Key Laboratory of Dark Matter and Space Astronomy, Purple Mountain
Observatory, Chinese Academy of Sciences, Nanjing 210008, P.R.China \\
$^b$School of Astronomy and Space Science, University of Science and
Technology of China, Hefei 230026, P.R.China
}

\begin{abstract}
Complicated hardenings and softenings of the spectra of cosmic ray protons 
and helium have been revealed by the newest measurements, which indicate 
the existence of multiple source populations of Galactic cosmic rays. 
We study the physical implications of these results in this work. 
A phenomenological fit shows that three components can properly give 
the measured structures of the proton and helium spectra. The data are 
then accounted for in a physically motivated, spatially-dependent 
propagation model. It has been shown that one background source population 
plus two local sources, or two background source populations plus one 
local source can well reproduce the measurements. The spectral structures 
of individual species of cosmic rays are thus natural imprints of different 
source components of cosmic rays. Combined with ultra-high-energy 
$\gamma$-ray observations of various types of sources, the mystery about 
the origin of Galactic cosmic rays may be uncovered in future.
\end{abstract}

\date{\today}

%95.35.+d: Dark matter
%96.50.S-: Cosmic rays

\maketitle

\section{Introduction}
The origin of cosmic rays (CRs) is a century-long unresolved problem in 
astrophysics. Precise measurements of the spectra of individual species 
are crucial to uncovering the origin of these energetic particles.
In recent years, direct measurements by space detectors have measured the 
energy spectra of a number of species up to 100 TeV with high precision 
\cite{2011Sci...332...69A,2015PhRvL.114q1103A,2015PhRvL.115u1101A,
2017PhRvL.119y1101A,2020PhRvL.124u1102A,2021PhRvL.126d1104A,
2018JETPL.108....5A,2019SciA....5.3793A,2024PhRvD.109l1101A,
2021PhRvL.126t1102A,2022PhRvL.129j1102A,2023PhRvL.130q1002A,
2025arXiv251105409D}. 
Spectral hardenings around hundreds of GV and softenings around 15 TV 
have been revealed, giving a bump-like structure in the spectra. Most 
recently, the Large High Altitude Air Shower Observatory (LHAASO) 
experiment measured the proton and helium spectra in the PeV region
with very high precision \cite{2025SciBu..70.4173C,2026PhRvL.136l1001C}. 
Compared with previous measurements
\cite{2013APh....47...54A,2019PhRvD.100h2002A}, the systematic 
uncertainties of the LHAASO results are well controlled, and the 
charge-dependent ``knees'' of these particle are shown clearly for 
the first time \cite{2025SciBu..70.4173C,2026PhRvL.136l1001C}. The LHAASO
spectra connect smoothly with the direct measurements by DAMPE, but their 
spectral difference suggest the existence of hardenings around 100 TV 
(see also the measurement by GRAPES-3 \cite{2024PhRvL.132e1002V}). 
Combined with the direct measurement results, the wide-band spectra of 
protons and helium from sub-GeV to 10 PeV show complicated structures 
as three bumps: one around several GV, one around 15 TV, and one around 
several PV. These spectral structures deviate from the expectation of 
the conventional acceleration and propagation theories which predict
power-law spectra of CRs at high energies, and lead to challenges
(also opportunities) to the understanding of CR physics.

The turnover at $\sim$GV is expected to be due to multiple reasons,
including the ionization and Coulomb energy losses, the transition from 
relativistic regime to non-relativistic regime, the solar modulation, 
and the possible break of the injection spectrum at production. 
The spectral hardening around hundreds of GV has attracted great 
attention of theoretical efforts. While many models have been proposed 
to account for this feature (e.g., \cite{2011ApJ...729L..13O,
2011PhRvD..84d3002Y,2012ApJ...752...68V,2012PhRvL.109f1101B,
2012ApJ...752L..13T,2012MNRAS.421.1209T,2013ApJ...763...47P}),
few of them can simultaneously reproduce the subsequent softening
above $\sim$15 TV. This joint bump structure (hardening-softening)
can be properly explained via introducing a local CR source
\cite{2015ApJ...809L..23S,2019JCAP...10..010L,2020ApJ...903...69F}, 
re-acceleration by a local shock \cite{2021ApJ...911..151M,
2022ApJ...933...78M,2026arXiv260214196H}, or the propagation in the 
environment with excitation/damping of magnetohydrodynamic waves 
\cite{2022ApJ...937..107C}. Given proper location of the local source, 
the energy evolution of the large-scale anisotropies can also be 
explained, which offers additional support of the local source model
\cite{2019JCAP...10..010L,2023ApJ...942...13Q}.
There were also many models to explain the ``knee'' of the spectra 
around several PV (see the compilation of \cite{2004APh....21..241H}), 
which can be classified into two categories: the acceleration or 
propagation category ($Z$-dependent ``knees'') and the interaction 
category ($A$-dependent ``knees''). The LHAASO results support the former
category. The joint spectra of direct and indirect measurements show an 
additional bump structure (hardening-softening) from 100 TeV to 10 PeV.
The apparently complicated structures of the energy spectra in a wide
energy range requires an overall explanation of these structures
simultaneously.

In this work we revisit the interpretations of these spectral structures.
The basic assumption is that there are multiple source populations
\cite{2013FrPhy...8..748G,2024A&A...692A..20R,2025ApJ...979..225L,
2024arXiv240313482Y,2025PhRvD.112l3033P}, which is natural according to 
observations of $\gamma$ rays \cite{2023ARNPS..73..341C}. The spectral 
structures can then be attributed to transitions among different source 
populations. The differences between proton and helium spectra are due to 
the diverse properties of the accelerated spectra and abundances for 
different source populations. We first fit the data using a phenomenological 
approach assuming simple spectral forms in Sec. II, and then in Sec. III we 
try to reproduce the data in a physical propagation framework with more 
realistic model settings. We conclude this work in Sec. IV.

\section{Phenomenological three-component fitting of the spectra}

To reproduce the observational two-bump structures of the proton and
helium spectra above tens of GeV, we introduce a three-component model. 
For each component, the spectrum is described as an exponentially cutoff 
power-law function of rigidity
\begin{equation}
\phi_{ij}({\mathcal R}) \propto {\mathcal R}^{-\alpha_{ij}}
\exp(-{\mathcal R}/{\mathcal R}^{ij}_c),
\end{equation}
where $i={\rm p~or~He}$, $j$ represents one of the three components,
$\alpha_{ij}$ is the spectral index, ${\mathcal R}^{ij}_c$ is the cutoff 
rigidity. We further require that the difference of spectral indices 
between protons and helium nuclei for the same component does not 
exceed 0.2. This requirement reflects similar properties of protons 
and helium for the same component, but enables slight difference which 
may be due to the acceleration or propagation processes.

\begin{figure}[!htb]
\includegraphics[width=0.48\textwidth]{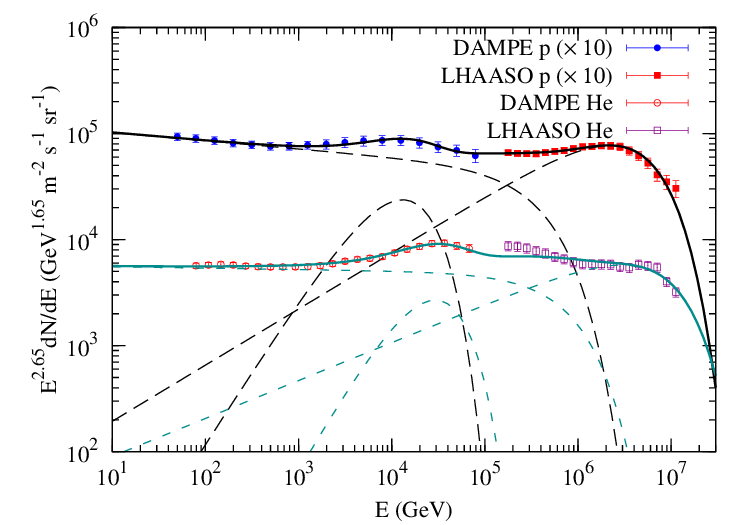}
\caption{Joint fiting results of the proton and helium spectra assuming 
three source components, compared with the measurement of DAMPE 
(FTFP\_BERT model) \cite{2019SciA....5.3793A,2021PhRvL.126t1102A} and 
LHAASO (EPOS-LHC model) \cite{2025SciBu..70.4173C,2026PhRvL.136l1001C}.}
\label{fig:3comp}
\end{figure}

The fitting results of the spectra are shown in Fig.~\ref{fig:3comp},
compared with the DAMPE (for FTFP\_BERT model) 
\cite{2019SciA....5.3793A,2021PhRvL.126t1102A} and LHAASO 
(for EPOS-LHC model) \cite{2025SciBu..70.4173C,2026PhRvL.136l1001C} data.
The obtained parameter values are given in Table \ref{table:3comp}.
Note that during the fitting, the systematic uncertainties associated
with the analysis of the measurements are added in quadrature to the 
statistical ones, and the hadronic model associated systematic 
uncertainties are not included. 

\begin{table}[!htb]
\centering
\caption {Fitting parameters of the three source components.}
\begin{tabular}{cccc}
\hline \hline
Component & $\alpha_{\rm p}$ & $\alpha_{\rm He}-\alpha_{\rm p}$ & ${\mathcal R}_c$\\
 &  &  & (PV) \\
\hline
1 & $2.74\pm0.02$ & $-0.07^{+0.02}_{-0.02}$ & $0.41^{+0.12}_{-0.09}$ \\
2 & $1.26\pm0.29$ & $-0.12^{+0.02}_{-0.08}$ & $0.010^{+0.003}_{-0.002}$ \\
3 & $2.18\pm0.07$ & $0.15^{+0.05}_{-0.01}$ & $4.55^{+0.49}_{-0.44}$ \\
\hline
\hline
\end{tabular}
\label{table:3comp}
\end{table}

The three components contribute to different features of the spectra. 
Component 1 has a relatively soft spectral shape and a sub-PV cutoff, 
dominating the low-energy part of the spectra. The spectrum of component 
2 is very hard, with a cutoff of $\sim0.01$ PV, which mainly contributes 
to the $O(10)$ TV bump. The spectrum of component 3 is harder than 
component 1, and the cutoff rigidity is about 4 PV, giving rise to the 
knee of the spectra. One can also note that the helium spectral index 
tends to be harder than that of protons for component 1 and 2, but is 
softer instead for component 3. Their spectral differences are necessary
to reproduce the apparently different spectra of protons and helium.
But note that the quantitative parameters depend on the assumed spectral
forms of Eq. (1).

Interestingly, the three source component may be related with different
source populations as revealed by ultra-high-energy $\gamma$-ray 
observations. Component 1 can be attributed to supernova remnants (SNRs), 
which have been shown can accelerate protons to sub-PeV energies 
\cite{2024SciBu..69.2833C}. Component 2 is very likely to be due to a 
local source, presumably an SNR such as that associated with Geminga pulsar 
\cite{2019JCAP...10..010L,2022ApJ...926...41Z}. The hard spectrum could
be naturally explained as low-energy particles do not diffuse from the 
source to the Earth within the lifetime of the source. See Sec. III for 
more details about this possibility. Component 3 could be from other 
types of energetic particle accelerators in the Milky Way, e.g., pulsar 
wind nebulae \cite{2003A&A...402..827A,2021Sci...373..425L}, 
massive star clusters \cite{2019NatAs...3..561A,2024SciBu..69..449L}, 
or black hole jets \cite{2024Natur.634..557A,2025NSRev..12af496L,
2026ApJ...997..163Y,2025arXiv251001369K,2025PhRvD.112l3015Z}.

\begin{figure*}[!htb]
\includegraphics[width=0.8\textwidth]{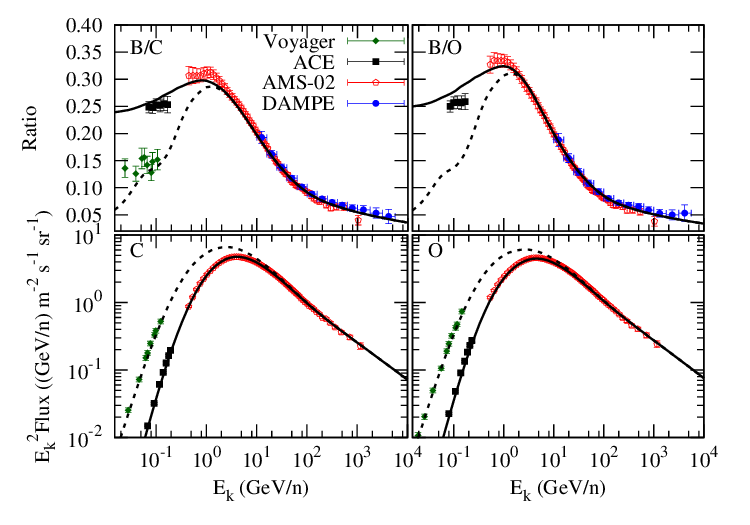}
\caption{Fitting results of the ratios of B/C, B/O, and fluxes of carbon
and oxygen in the SDP model, compared with the measurement of Voyager-1
\cite{2016ApJ...831...18C}, ACE \cite{2019SCPMA..6249511Y}, AMS-02 
\cite{2021PhR...894....1A}, and DAMPE \cite{2022SciBu..67.2162D}.
Dashed lines are the results in the local interstellar medium, and solid 
lines are those after solar modulation with a potential of $600$ MV.}
\label{fig:bco_sdp}
\end{figure*}

\section{Modeling of the spectra in a physical propagation framework}

\subsection{Propagation model}

The discussion in the previous section is empirical. In this section,
we try to address the measured spectral features in a more physical
origin and propagation model of Galactic CRs. The propagation of
Galactic CRs is characterized by a diffusion process, with possible
convection and reacceleration effects \cite{2007ARNPS..57..285S}.
The geometry of the Milky Way halo is usually approximated as a
cylinder, with radial boundary $r_h$ and vertical height $\pm z_h$.
The diffusion coefficient is usually parameterized as a power-law 
form of rigidity, $D_{xx}({\mathcal R})=D_0\beta^{\eta}({\mathcal R}/
{\mathcal R}_0)^{\delta}$, where $\beta$ is the particle velocity 
in unit of light speed, $\delta$ is the index related with the 
properties of the interstellar turbulence, and $\eta$ is employed 
to better reproduce the low-energy data of secondary-to-primary ratios. 
In this work we employ the {\tt GALPROP} 
package\footnote{https://galprop.stanford.edu/} 
\cite{1998ApJ...509..212S,2022ApJS..262...30P} to calculate the 
propagation of CRs. The diffusion reacceleration model is adopted
\cite{2011ApJ...729..106T,2020JCAP...11..027Y}, and the Alfven speed
$v_A$ is adopted to characterize the reacceleration effect
\cite{1994ApJ...431..705S}.

Recent observations of TeV halos around pulsars 
\cite{2017Sci...358..911A,2021PhRvL.126x1103A} suggest that the 
diffusion coefficient around the pulsars is much smaller than the 
average one inferred from CR secondary-to-primary ratios, indicating 
that the diffusion should be spatially-dependent 
\cite{2018ApJ...863...30F,2018PhRvD..97l3008P}. It has also been 
proposed that the spatially-dependent propagation (SDP) model can 
account for several observational anomalies of CRs and $\gamma$ rays 
\cite{2012ApJ...752L..13T,2012PhRvL.108u1102E,2016MNRAS.462L..88R,
2018PhRvD..97f3008G,2022FrPhy..1744501Q}.
In this work we adopt the two-halo approach\footnote{In reality the slow 
diffusion regions may not occupy the full Galactic disk, and the spatial 
distribution of the diffusion coefficient may be very complicated
\cite{2019ApJ...879...91J}. We use the two-halo model to effectively
describe the slow diffusion in the vicinity of sources and fast
diffusion otherwise, with proper filling factor of the slow diffusion
regions \cite{2024PhRvD.110j3039S}.} of the diffusion coefficient,
with slow diffusion in a thin disk and a fast diffusion in a thick halo.
Note that if one only discusses the spectral structures of CRs, 
the SDP scenario is not necessarily required. Nevertheless, the basic 
conclusion of this work, i.e., the requirement of multiple source 
components, is not affected by such a setting of the SDP model.
The diffusion coefficient is parameterized as \cite{2023FrPhy..1844301M}
\begin{equation}
D_{xx}({\mathcal R},z)=f(z)D_0\beta^{\eta}\left(\frac{{\mathcal R}}
{{\mathcal R}_0}\right)^{\delta_1}\left[1+\left(\frac{\mathcal R}
{{\mathcal R}_1}\right)^2\right]^{(\delta_2-\delta_1)/2},
\end{equation}
with ${\mathcal R}_0\equiv 4$ GV, $f(z)=\xi+(1-\xi)[1-\exp(-z^2/2h^2)]$, 
where $h$ is the characteristic thickness of the slow-diffusion disk, 
$\xi$ is a constant describing the suppression factors of $D_0$ in the 
disk region with respect to the halo region. This parameterization form 
connects the disk and halo smoothly. Here we introduce a break of the 
rigidity dependence of the diffusion coefficient, with slopes $\delta_1$ 
and $\delta_2$ below and above ${\mathcal R}_1$, in both the disk and 
halo regions to account for the observational breaks of B/C and B/O
ratios \cite{2022SciBu..67.2162D}. Different assumptions of abrupt
or gradually changing $\delta$ from the disk to the halo 
\cite{2012ApJ...752L..13T,2018PhRvD..97l3008P} are expected to result 
in similar effects on the observed spectra.

The propagation parameters are obtained through fitting to the newest
measurements of the B/C and B/O ratios, together with the C and O fluxes
from $\sim0.01$ GeV/n to $\sim5$ TeV/n by Voyager-1 
\cite{2016ApJ...831...18C}, ACE \cite{2019SCPMA..6249511Y},
AMS-02 \cite{2021PhR...894....1A}, and DAMPE \cite{2022SciBu..67.2162D}. 
See also Refs.~\cite{2016PhRvD..94l3007F,2021PhRvD.104l3001Z} for 
constraining the SDP parameters with different parameterizations using 
the Bayesian inference. Note that we use the same time period as AMS-02 
from May, 2011 to May, 2018 to extract the ACE data to enable a consistent 
solar modulation effect \cite{2019SCPMA..6249511Y}. The solar modulation 
is described with the force-field approximation \cite{1968ApJ...154.1011G}. 
The injection spectrum of primary C and O is assumed to be a smoothly 
broken power law form of rigidity
\begin{equation}
q({\mathcal R})=q_0{\mathcal R}^{-\gamma_1}\left[1+\left(\frac
{\mathcal R}{{\mathcal R}_{\rm br}}\right)^2\right]^{(\gamma_1-\gamma_2)/2},
\label{eq:inj}
\end{equation}
where $\gamma_1$ and $\gamma_2$ are the spectral indices below and above
the break rigidity ${\mathcal R_{\rm br}}$. The break of the injection
spectrum occurs at relatively low rigidity to better reproduce the
low-energy spectra. 

The main propagation parameters we obtain are: $z_h=4.0$ kpc, 
$D_0=4.5\times10^{28}$ cm$^2$~s$^{-1}$, $\delta_1=0.48$, $\delta_2=0.20$, 
${\mathcal R}_1=250$ GV, $\eta=-1.0$, $v_A=15$ km~s$^{-1}$, $h=0.45$ kpc, 
$\xi=0.1$. Fitting results of the B/C, B/O, C, and O fluxes, together with 
the measurements are shown in Fig.~\ref{fig:bco_sdp}. The solar modulation 
is described by the force-field approximation \cite{1968ApJ...154.1011G}, 
with a modulation potential of $\sim600$ MV. One can see that the 
observational spectral hardenings of both primary spectra and 
secondary-to-primary ratios can be reproduced in this model.

\begin{figure*}[!htb]
\includegraphics[width=0.48\textwidth]{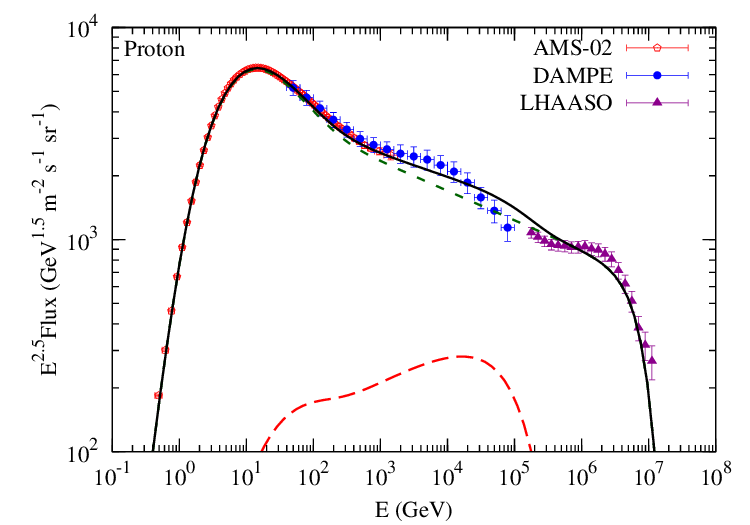}
\includegraphics[width=0.48\textwidth]{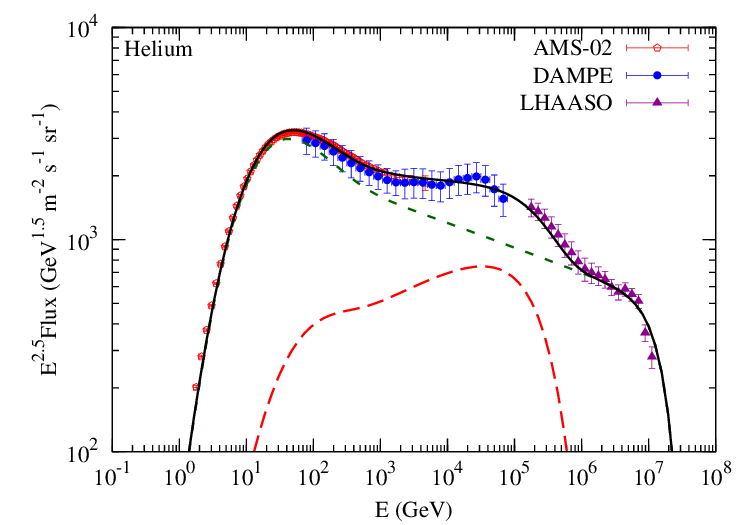}
\caption{Model results of the proton and helium spectra assuming the 
contribution from background sources (green short-dashed) and a local
source (red long-dashed). Solid lines show the total fluxes. The solar
modulation potential is 600 MV.}
\label{fig:phe_sdp}
\end{figure*}

\begin{figure*}[!htb]
\includegraphics[width=0.48\textwidth]{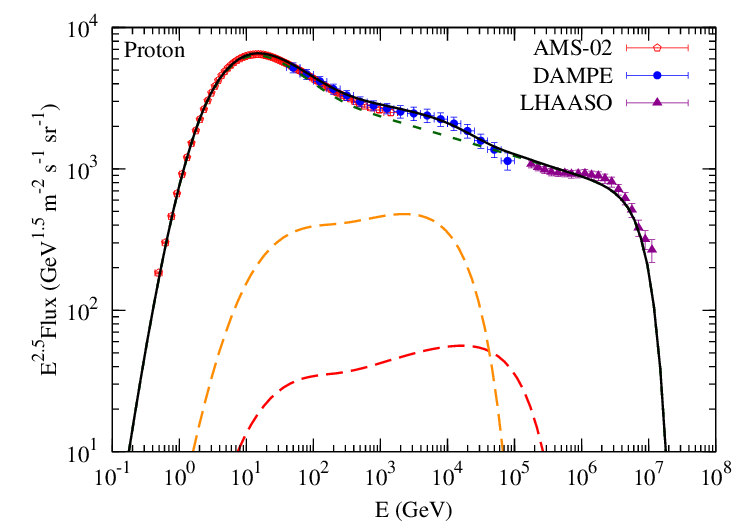}
\includegraphics[width=0.48\textwidth]{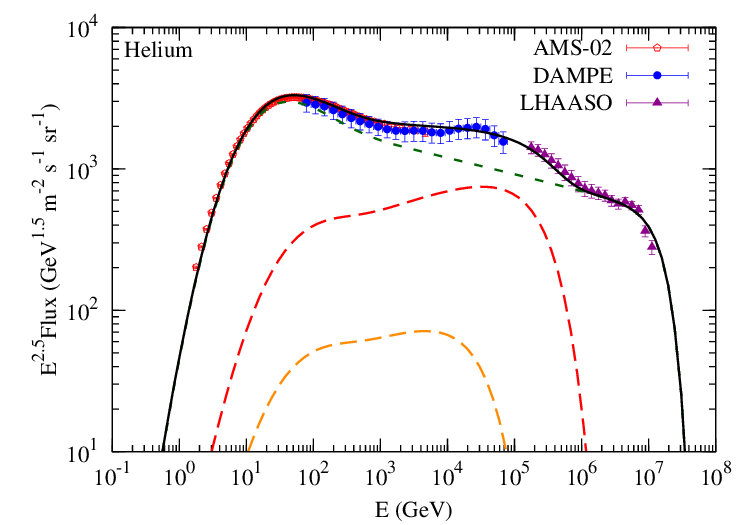}
\caption{Model results of the proton and helium spectra assuming the 
contribution from background sources (green short-dashed) and two local
sources (long-dashed). Solid lines show the total fluxes. The solar
modulation potential is 600 MV.}
\label{fig:phe_sdp2}
\end{figure*}

\subsection{Local source}

The complicated structures of the proton and helium spectra are hard to 
be reproduced purely by the propagation model without finely tuning the 
injection spectrum. As indicated by the pattern change of the large-scale 
anisotropies around $O(100)$ TeV \cite{2017ApJ...836..153A,
2017PrPNP..94..184A}, a local source component may naturally account for
these structures \cite{2015ApJ...809L..23S,2019JCAP...10..010L,
2020ApJ...903...69F}. The injection spectrum of the local source is 
described by Eq. (1). Since we consider mainly the impact of the local 
source on the spectra above 100 GeV, a spherically symmetric plain 
diffusion model is adopted to calculate the propagation of CRs from the 
local source. The diffusion coefficient is used as the disk value of the 
SDP scenario. The solution of the propagated spectrum, assuming a 
burst-like injection from a point source, is
\begin{equation}
\phi(r,t,{\cal R})=\frac{q({\cal R})}{(\sqrt{2\pi}\sigma)^3}
\exp\left(-\frac{r^2}{2\sigma^2}\right),
\end{equation}
where $r$ is the distance from the source, $t$ is the time from the 
injection, $q({\cal R})$ is the injection spectrum, and
$\sigma({\cal R},t)=[2D_{xx}({\cal R})t]^{1/2}$ is the effective diffusion 
length within time $t$.

\subsection{Results}

Now we investigate the proton and helium spectra in the SDP model.
The measurements in a very wide energy range from $\sim$GeV to $\sim10$
PeV by AMS-02 \cite{2021PhR...894....1A}, 
DAMPE \cite{2019SciA....5.3793A,2021PhRvL.126t1102A}, and 
LHAASO \cite{2025SciBu..70.4173C,2026PhRvL.136l1001C} are used. 
The injection spectrum of protons and helium are also described by 
Eq. (3). However, to reproduce the ``knees'' of the spectra, 
an additional super-exponential cutoff term of
$\exp\left[-\left({\mathcal R}/{\mathcal R}_c^{\rm bkg}\right)^2\right]$
is multiplied to Eq. (3). We find that ${\mathcal R}_c^{\rm bkg}=9$ PV 
is proper to account for the LHAASO data. Other parameters of the
injection spectra of protons and helium are given in Table
\ref{table:bkg}.

For the local source, we take the Geminga SNR as an example 
\cite{2022ApJ...926...41Z}. The distance is $r=0.25$ kpc 
\cite{1994A&A...281L..41S}), and the age is $t=3.4\times10^5$ yr 
\cite{2005AJ....129.1993M}. For spectral index $\alpha_{\rm loc}=2.1$ 
and cutoff rigidity ${\mathcal R}_c^{\rm loc}=120$ TV, we show the obtained 
results of the proton and helium spectra in Fig.~\ref{fig:phe_sdp}. 
The match between the model and data is, however, 
not good, due to that the required local source contribution of helium 
appears at higher energies than that of protons rather than the simple 
charge scaling based on the same rigidity. This result is also consistent 
with that obtained in Sec. II, where different spectral indices of protons 
and helium for components 1 and 3 are employed to resolve this issue.

\begin{table}[!htb]
\centering
\caption{Spectral parameters of background component.}
\begin{tabular}{ccccc}
\hline \hline
 & $\gamma_1$ & $\gamma_2$ & ${\mathcal R}_{\rm br}$ & ${\mathcal R}_c^{\rm bkg}$ \\
 &  &  & (GV) & (PV) \\
\hline
p  & $2.05$ & $2.44$ & $11.0$ & $9.0$ \\
He & $2.20$ & $2.42$ & $11.0$ & $9.0$ \\
\hline
p (I)  & $2.00$ & $2.53$ & $11.0$ & $0.3$ \\
He (I) & $2.18$ & $2.36$ & $11.0$ & $0.3$ \\
p (II)  & ... & $2.30$ & ... & $6.5$ \\
He (II) & ... & $2.36$ & ... & $6.5$ \\
\hline
\hline
\end{tabular}
\label{table:bkg}
\end{table}

\begin{table}[!htb]
\centering
\caption {Spectral index, cutoff rigidity, kinetic energies of protons and 
helium nuclei above 1 GeV of the local source(s).}
\begin{tabular}{ccccc}
\hline \hline
 & $\alpha_{\rm loc}$ & ${\mathcal R}_c^{\rm loc}$ & $W_{\rm p}$ ($>1$~GeV) & $W_{\rm He}$ ($>1$~GeV) \\
 &  & (TV) & ($10^{50}$~erg) & ($10^{50}$~erg) \\
\hline
1 src (1 bkg) & $2.10$ & $120$ & $1.0$ & $1.8$ \\
\hline
2 src (1 bkg)     & $2.10$ & $15$ & $2.0$ & $0.2$ \\
  & $2.10$ & $120$ & $0.2$ & $1.8$ \\
\hline
1 src (2 bkg) & $2.10$ & $25$ & $2.5$ & $0.7$ \\
\hline
\hline
\end{tabular}
\label{table:src}
\end{table}

\begin{figure*}[!htb]
\includegraphics[width=0.48\textwidth]{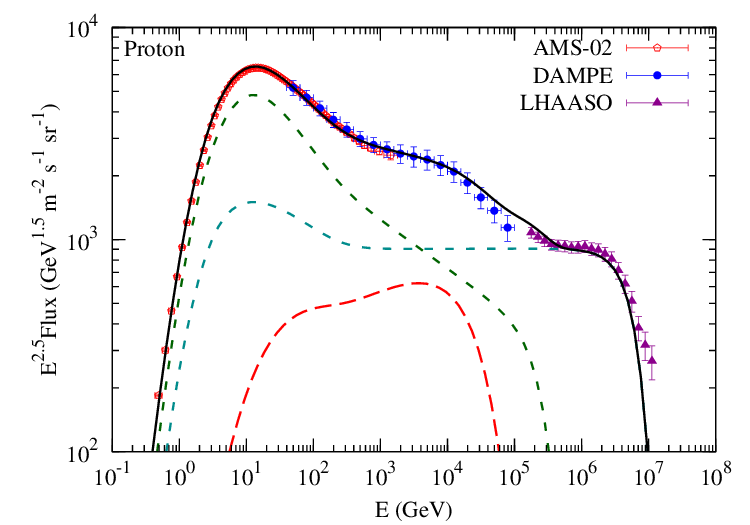}
\includegraphics[width=0.48\textwidth]{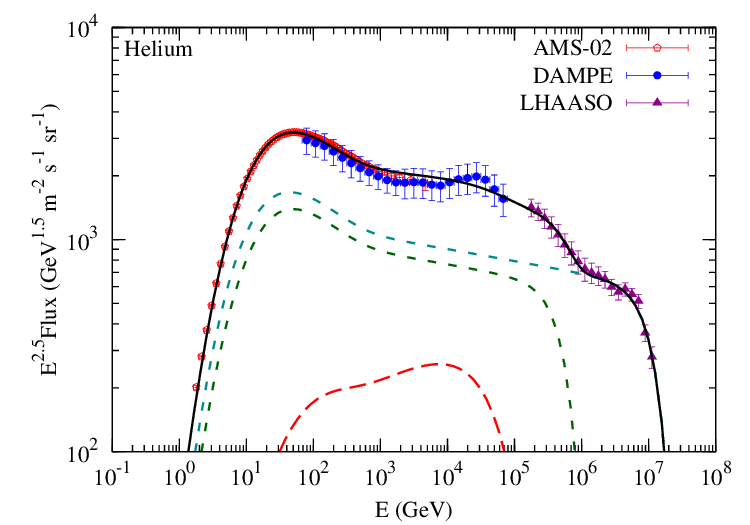}
\caption{Model results of the proton and helium spectra assuming the 
contribution from two populations of background sources (short-dashed) 
and one local source (red long-dashed). Solid lines show the total fluxes.
The solar modulation potential is 600 MV.}
\label{fig:phe_sdp3}
\end{figure*}

We propose that adding another local source with different proton-to-helium
abundance can reconcile this discrepancy. For simplicity, the distances and
ages of both sources are assumed to be the same, and the spectral indices 
are fixed to be 2.10. Source 1 has a relatively low cutoff rigidity of $15$ 
TV and a higher proton abundance, which mainly contributes to the proton 
bump, while source 2 has a higher cutoff rigidity of $120$ TV and a higher 
helium abundance to account for the helium bump. Comparison between
the model predictions and the data are presented in Fig.~\ref{fig:phe_sdp2}. 
The parameters of the local sources are given in Table \ref{table:src}. 
The observed proton and helium spectra can be properly reproduced in this
scenario. The kinetic energies of CRs are on the order of $10^{50}$ erg,
which is reasonable for typical remnants of supernova explosions.
The abundances of protons and helium are quite different for these two 
sources, as can be seen in Table \ref{table:src}. It is not common that 
helium is more abundant than protons in a CR source. Nevertheless, the 
abundance depends on the energy spectrum of particles. If the spectrum
of helium is harder than that of protons, the required helium energy can
be lowered down.

Another possibility is that there are more than one background source
populations, as discussed in Sec. II (see also \cite{2026ApJ...997..163Y}). 
Assuming two populations of background sources with different spectral 
parameters, together with one local source, the observational data can 
also be accounted for, as shown in Fig.~\ref{fig:phe_sdp3}. The local 
source parameters are given in Table~\ref{table:src}, and the background 
source parameters are given in Table~\ref{table:bkg}. To minimize the 
number of free parameters, a single power-law form with super-exponential 
cutoff form for background source population II is assumed. We can see 
that the high-energy helium spectral index is harder than that of protons 
for population I, and is softer for population II, just as the case
discussed in Sec. II. Since the observational proton and helium 
spectra are quantitatively different, it is difficult to reproduce the 
measurements within the three-component framework assuming the same 
spectral shapes of all the species and all components. Adding more source 
components with different proton and helium abundance ratios may solve this 
issue. Alternatively, the different spectral shapes of different particle 
species and different source components might be due to environment effect 
of acceleration by different sources \cite{2011ApJ...729L..13O}.
In Ref.~\cite{2026ApJ...997..163Y} it was proposed that population I may 
correspond to SNRs and population II may correspond to microquasars.

\section{Conclusion and discussion}

The newest measurements of the individual spectra of CR protons and helium 
nuclei up to 10 PeV by LHAASO show clearly the ``knees'' of those species 
as expected long time ago \cite{2025SciBu..70.4173C,2026PhRvL.136l1001C}.
Together with the direct measurements below $\sim 100$ TeV
\cite{2021PhR...894....1A,2019SciA....5.3793A,2021PhRvL.126t1102A}, 
the detailed spectral structures in a very wide energy band have been 
revealed for the first time, which offer very useful insights in
understanding the origin of Galactic CRs. We study possible implications 
on the sources of Galactic CRs based on these new measurements.

The spectra of both protons and helium show two-bump structures above
tens of GeV. However, there apparent features are different, especially
for energies above 100 TeV. A phenomenological fitting of the spectra,
assuming simple exponentially cutoff power-law rigidity spectrum,
indicates that three source components are necessary to account for
the data. The spectral indices and relative abundances of protons
and helium are not identical among the three components, in order to 
give the apparently different spectral shapes of protons and helium.
The three source components may correspond to two different source
populations with different maximum acceleration limits, together with
a local source. The ultrahigh energy $\gamma$-ray observations of
various types of Galactic sources may give hints to these source
populations.

We then explain the measured spectra in a physical propagation model. 
Particularly, we consider the SDP framework as implied by recent
observations of $\gamma$ rays and secondary CRs. We ascribe the majority
of the broadband spectra from sub-GeV to 10 PeV to a single background
source population, and the $\sim15$ TV bump to a local source. This
setting is found not enough to explain the data well due to the different
spectra behaviors of protons and helium nuclei. Via adding another local
source with different nuclear abundance, the measurements can be well
accounted for. Alternatively, if only one local source is introduced,
two background source populations as inferred in the phenomenological 
fitting with different spectral shapes of protons and helium can also
explain the data. 

The existence of one or more local source(s) is natural given that a 
number of high-energy CR accelerators within a few hundred pc have been 
observed in $\gamma$ rays. The local source(s) can also explain the 
observed phase flip and amplitude spectrum of large-scale anisotropies
\cite{2019JCAP...10..010L}. It has been proposed that the remnant
of supernova explosion associated with Geminga pulsar could be a good
candidate of the local source, given its proper direction, distance,
and age \cite{2022ApJ...926...41Z}. 
Given its proximity, the nearby source may leave imprints on the 
electron spectrum and diffuse $\gamma$ rays. With the parameters adopted
in this work, we check that the expected contribution to the electron
spectrum and diffuse $\gamma$-ray spectrum is consistent with the 
current measurements \cite{2014PhRvL.113v1102A,2017Natur.552...63D,
2023PhRvL.131s1001A,2015ApJ...799...86A} (see the Appendix for more details). 
Particularly, since the Earth is surrounded by CR particles from 
the local source, the $\gamma$-ray surface brightness profile is expected 
to be rather flat, and the emission is highly diffuse (Fig.~\ref{fig:g} in
the Appendix). The detection of such a diffuse and relatively weak 
$\gamma$-ray component may be challenging. 
Furthermore, the electrons and positrons produced by the pulsar or its 
nebula associated with the SNR may contribute to solving the puzzle of 
$e^+e^-$ excesses \cite{2021JCAP...05..012Z}. 
Very interestingly, the contribution to electrons from the SNR may 
be required to interpret simultaneously the positron flux, whose dropoff 
at $\sim284$ GeV by AMS-02 \cite{2019PhRvL.122d1102A} might suggest the 
existence of a bump excess of the electron component 
\cite{2021JCAP...05..012Z}.
The $\gamma$-ray emission from local giant molecular clouds is expected 
to be able to critically test the local source scenario 
\cite{2025PhRvD.111j3017L}. 

Multiple background source populations are also expected in the Milky Way.
The $\gamma$-ray observations show that several types of sources can
accelerate CRs to $\sim$PeV energies, such as SNRs, pulsar wind nebulae, 
massive star clusters, and microquasars \cite{2023ARNPS..73..341C}. 
The overall properties of the accelerated CR particles, such as the spectral 
indices and abundances of different species, and the acceleration limits, 
need further observational and theoretical studies of these objects. We 
emphasize that, apart from spectra of CRs, the joint analysis of correlated 
structures on the composition and anisotropies would provide critical 
insights in understanding the origin of CRs \cite{2025arXiv250618118Q}.

The data that support the findings of this article are openly
available \cite{Note}.

\acknowledgments
This work is supported by the National Natural Science Foundation of China 
(No. 12220101003) and the Project for Young Scientists in Basic Research 
of Chinese Academy of Sciences (No. YSBR-061).

\setcounter{figure}{0}
\renewcommand\thefigure{A\arabic{figure}}

\appendix*
\section{Contribution to the electron and gamma-ray fluxes from the 
nearby source}

\begin{figure}[!htb]
\includegraphics[width=0.48\textwidth]{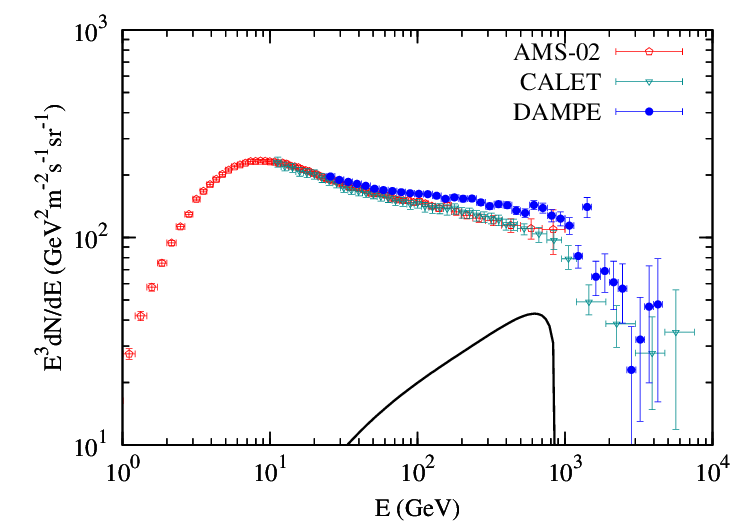}
\caption{The electron flux from the local source, compared with the 
measurements \cite{2014PhRvL.113v1102A,2017Natur.552...63D,
2023PhRvL.131s1001A}.}
\label{fig:e}
\end{figure}

Fig.~\ref{fig:e} shows the expected flux of electrons from the
local source, assuming the same injection parameters as protons 
($\alpha_{\rm loc}^e=2.1$, ${\mathcal R}_c^{\rm loc}=25$ TV) but with
an energy budget of $1.3\times10^{48}$ erg (corresponding to an 
electron-to-proton ratio of $\sim5\times10^{-3}$). The theoretical flux 
is lower than the measurements (for total electrons plus positrons)
\cite{2014PhRvL.113v1102A,2017Natur.552...63D,2023PhRvL.131s1001A},
indicating that the model setting of this work is self-consistent.

\begin{figure}[!htb]
\includegraphics[width=0.47\textwidth]{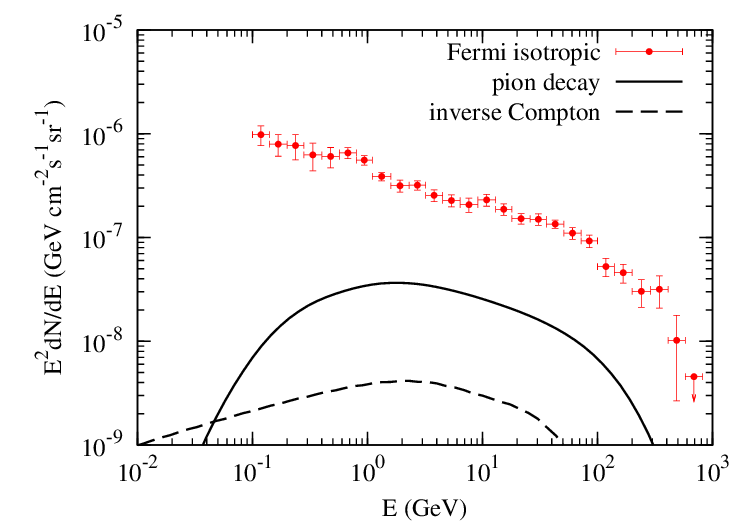}
\includegraphics[width=0.48\textwidth]{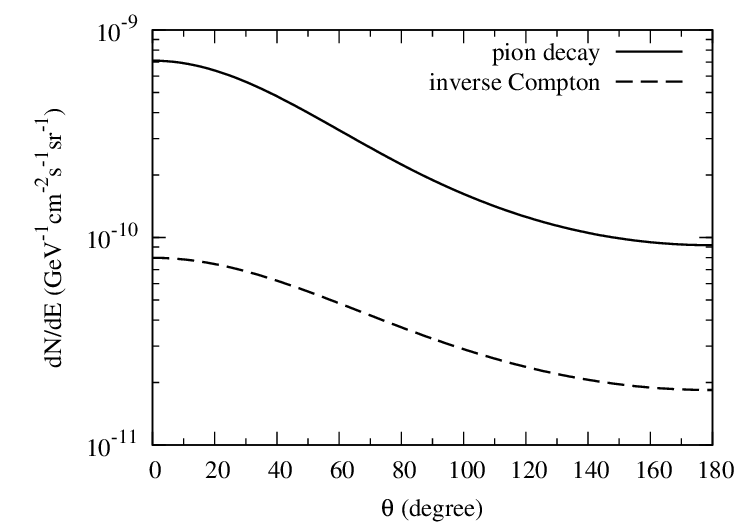}
\caption{Top: the all-sky diffuse $\gamma$-ray flux from the local source, 
compared with the extragalactic $\gamma$-ray background measured by
Fermi-LAT \cite{2015ApJ...799...86A}.
Bottom: the surface brightness profile of $\gamma$-ray flux from the 
local source at energy of 10 GeV.}
\label{fig:g}
\end{figure}

Fig.~\ref{fig:g} shows the expected all-sky diffuse $\gamma$-ray spectrum
(top panel) and the surface brightness profile at 10 GeV (bottom panel)
from the local source (with $\alpha_{\rm loc}=2.1$, ${\mathcal R}_c^{\rm loc}
=25$ TV, $W_p=2.5\times10^{50}$ erg, and $W_e=1.3\times10^{48}$ erg), 
for the neutral pion decay and the inverse Compton scattering components, 
respectively. For simplicity, the number density of the interstellar medium 
is assumed to be 1 hydrogen cm$^{-3}$, and the interstellar radiation field 
is approximated with three gray-body components: the cosmic microwave 
background with temperature of 2.725 K and energy density of $0.26$ eV 
cm$^{-3}$, an infrared background with temperature of 20 K and energy 
density of $0.3$ eV cm$^{-3}$, and an optical background with temperature 
of 5000 K and energy density of $0.3$ eV cm$^{-3}$ \cite{2017Sci...358..911A}. 
The $\gamma$-ray emission is very extended compared with detectable
sources. As a reference, the maximum difference of the emission is about a
factor of 8 across the whole sky, while the surface brightness profile of
the Geminga halo detected by HAWC differs by a factor of about 1000 within
10 degrees \cite{2017Sci...358..911A}. Furthermore, the flux of the diffuse
emission is relatively weak. Compared with the Fermi-LAT measurement of the 
extragalactic $\gamma$-ray background \cite{2015ApJ...799...86A}, the 
expected flux from the local source is lower by about an order of magnitude. 
If one considers the diffuse emission from the Galactic plane
\cite{2012ApJ...750....3A}, the flux from the local source is even smaller 
by another order of magnitude. The detection of such a very extended, weak 
diffuse emission is thus challenging with the current experiments. 
Nevertheless, we can find that the current model setting is consistent with 
the $\gamma$-ray observations.

\bibliographystyle{apsrev}
\bibliography{/home/yuanq/work/material/refs}

\end{document}